\documentclass[10pt,a4paper,twoside]{article}
\usepackage{epsfig}
\usepackage{baltlat5}
\usepackage{wrapfig}
\begin{document}
\ \
\vspace{0.5mm}

\setcounter{page}{1}
\vspace{5mm}

\titlehead{Baltic Astronomy, vol.\ts 14, XXX--XXX, 2005.}

\titleb{TIME RESOLVED SPECTROSCOPY OF KPD 1930+2752}

\begin{authorl}
\authorb{S.~Geier}{1}
\authorb{U.~Heber}{1}
\authorb{N.~Przybilla}{1}
\authorb{R.-P.~Kudritzki}{2}
\end{authorl}

\begin{addressl}
\addressb{1}{Dr.-Remeis-Sternwarte, Universit\"at Erlangen-N\"urnberg, Sternwartstrasse 7, \\
96049 Bamberg, Germany}

\addressb{2}{Institute for Astronomy, University of Hawaii, 2680 Woodlawn Drive, \\
Honolulu, HI 96822, USA}

\end{addressl}

\submitb{Received 2005 April 1}

\begin{abstract}

We present the analysis of 200 high-resolution and 150 low-resolution spectra of the pulsating subdwarf close binary KPD 1930+2752 obtained with the Keck and the Calar Alto 2.2 m telescopes. Using metal-rich, line blanketed LTE model atmospheres the atmospheric parameters \(T_{\rm eff} = 35\,500 \pm 500 ~ {\rm K}\), \(\log \, g = 5.56 \pm 0.05\) and \(\log \, \frac{\rm He}{\rm H} = -1.48 \pm 0.02\) were derived. The radial velocity curve was measured and combined with all available published and unpublished radial velocity data  in order to get results of high accuracy (\(K = 341 \pm 1~{\rm km\,s^{-1}}\), \(P = 0.0950933 \pm 0.0000015~{\rm d}\)). For the first time we measured the projected rotational velocity. The preliminary result is \(v_{\rm rot}\sin{i} = 97 \pm 9~{\rm km\,s^{-1}}~(3\sigma )\).
Since the rotation of the sdB star is very likely tidally locked to the orbit, we can constrain the inclination of the system. Assuming the companion to be a white dwarf, the mass of the sdB is limited from \(0.44~M_{\odot}\) to \(0.63~M_{\odot}\) and the total mass range of the system is \(1.3~M_{\odot}\) to \(2.0~M_{\odot}\).
It is very likely that the total mass exceeds the Chandrasekhar limit. Hence KPD 1930+2752 is a candidate for a progenitor of a Type Ia supernova. According to the derived limits for the inclination angle, eclipses are likely to occur.

\end{abstract}

\begin{keywords}
binaries: spectroscopic --
stars: atmospheres --
stars: individual (KPD 1930+2752) --
subdwarfs --
supernovae: general
\end{keywords}
\newpage

\resthead{Time Resolved Spectroscopy of KPD 1930+2752}{S.~Geier, U.~Heber, N.~Przybilla, R.-P.~Kudritzki}

\sectionb{1}{INTRODUCTION}

KPD 1930+2752 was identified as a subdwarf B star in the Kitt Peak-Downes survey (Downes 1986). The parameters which were derived from spectroscopy by model atmosphere fits (Saffer and Liebert 1995) are consistent with the theoretical instability strip for pulsating sdB stars, which was predicted by Charpinet et al. (1996). After the first pulsating sdBs (EC 14026 stars) were discovered in 1997 (Kilkenny et al. 1997; Koen et al. 1997; Stobie et al. 1997), Bill\`{e}res et al. (2000) initiated a survey to search for these objects in the northern hemisphere. They selected KPD 1930+2752 from the list of Saffer and Liebert (1995) for their fast photometry program and detected multiperiodic variations with short periods and low amplitudes. In addition to 44 \pmode~pulsations they found a strong variation at a much longer period of about \(4100 \, {\rm s}\). This variation could be identified as an elipsoidal deformation of the sdB most likely caused by a massive companion. Bill\`{e}res et al. predicted the period of the binary to be two times the period of the brightness variation (\(P=8217.8 \, {\rm s} = 0.095111 \, {\rm d}\)).\\
This was proven by Maxted et al. (2000), who measured a radial velocity curve of KPD 1930+2752 which could be fitted with the proper period. The radial velocity amplitude \(K=349.3 \pm 2.7 \, {\rm km\,s^{-1}} \) combined with the assumption of the canonical mass for sdBs \(M_{ \rm sdB}=0.5 \, M_{\odot}\) led to a lower limit for the mass of the system derived from the mass function. This lower limit \(M \geq 1.47 M_{\odot} \) exceeded the Chandrasekhar mass of \(1.4 \, M_{\odot}\). Because there was no sign of a companion in the spectra, it was concluded that the unseen object must be a white dwarf. Putting all this together Maxted et al. concluded that KPD 1930+2752 could be a good candidate for the progenitor of a Type Ia supernova.\\
Type Ia supernovae (SNe Ia) are the most important standard candles for extragalactic distance measurements and play an outstanding role in observational cosmology. The progenitors of SNe Ia as well as the dynamics of the explosions are still under debate. The thermonuclear explosion of a white dwarf turned out to be the most reasonable explanation. A white dwarf has to exceed the Chandrasekhar mass to explode in this way. In the single degenerate scenario the white dwarf accretes matter from a nearby non-degenerate companion star. Possible progenitor candidates are supersoft X-ray sources. Double degenerate systems consist of two white dwarfs in a short period system, which merge within a Hubble time due to gravitational wave radiation (Livio 2000). KPD 1930+2752 has a sufficiently short period and may have a sufficiently high total mass. The sdB can be expected to evolve to a white dwarf before the merger.\\
From the theoretical point of view Ergma et al. (2001) questioned the double degenerate scenario in the case of KPD 1930+2752. Their simulations based on the derived parameters of Maxted et al. suggested the formation of a single massive ONeMg white dwarf.\\
The main drawback of all previous investigations is the lack of information on the inclination angle and the assumption made on the sdB mass. We drop the latter assumption and derive constraints on the inclination for the first time by means of an accurate measurement of the projected rotational velocity and surface gravity. Because the rotation of the sdB star is tidally locked to the orbit, we can derive \(\sin{i}\) as a function of the sdB mass from these two quantities.\\

\sectionb{2}{OBSERVATIONS AND DATA REDUCTION}

With the \(10 \, {\rm m}\) Keck Telescope at the Mauna Kea Observatory two hundred high-resolution spectra were obtained by N. Przybilla in half a night in July 2004, using the High Resolution Echelle Spectrometer (HIRES, Vogt et al. 1994). The spectra covered a wavelength range of \(4\,200 \, {\rm \AA} - 6\,800 \, {\rm \AA} \) with few small gaps at a resolution of \(0.1 \, {\rm \AA} \) and exposure times of \(20 \, {\rm s}\) each. The data were reduced using the ESO-MIDAS package. Bias and flatfield corrections were applied and a wavelength calibration was done. All spectra were corrected to the heliocentric frame of reference. \\
Additional observations were obtained with the \(2.2 \, {\rm m}\) Telescope at the Calar Alto Observatory in July 2004. The Calar Alto Faint Object Spectrograph (CAFOS) was used to obtain 150 spectra covering a wavelength range of \(3\,600 \, {\rm \AA} - 6\,200 \, {\rm \AA} \) with \(5 \, {\rm \AA} \) resolution and an exposure time of \(180 \, {\rm s}\) each. The data were reduced in analogy to the HIRES spectra.

\sectionb{3}{STELLAR PARAMETER DETERMINATION}

\begin{wrapfigure}[23]{i}[0pt]{72mm}
\psfig{figure=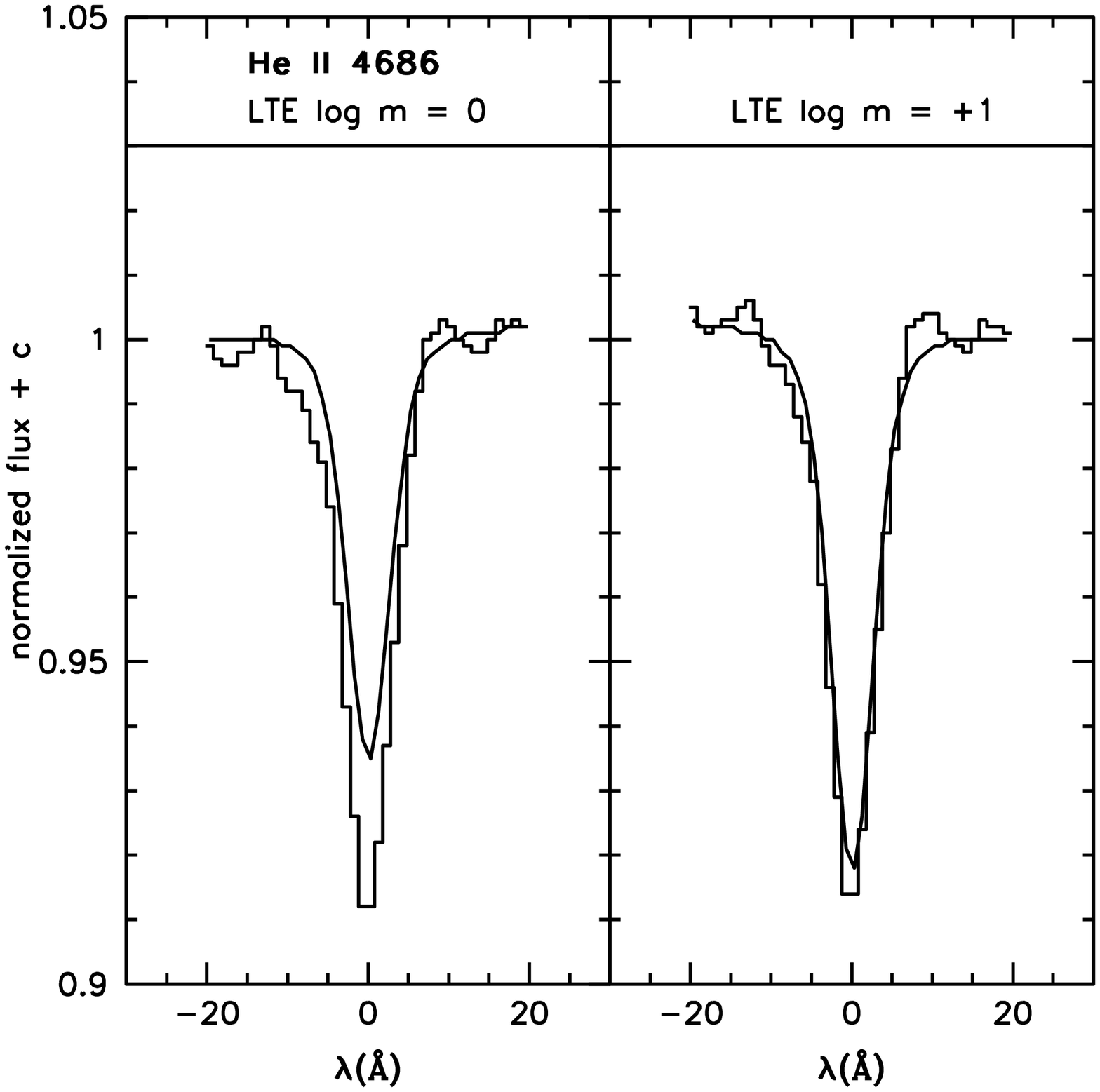,width=70mm,clip=}
\quote
\captionb{1}{LTE models with \(\log m=+1\) solve the helium problem.}
\end{wrapfigure}

The coadded CAFOS spectrum had a very smooth continuum, which made it very suitable for model atmosphere fits. Different LTE and NLTE model grids were used to fit 7 hydrogen and 8 He\,{\sc i} and He\,{\sc ii} lines: A grid of metal-line blanketed LTE model atmospheres (Heber et al. 2000) with solar and ten times solar metalicity; a grid of partially line-blanketed NLTE model atmospheres without metal lines (Dreizler et al. 1990) based on the ALI method (PRO2); a grid calculated with a new version of PRO2 (so called NGRT) that employs a temperature correction scheme and uses more sophisticated model atoms;  a small grid based on the LTE models with NLTE line formation (Przybilla et al. 2006). The matching of the observed spectra was done with the FITPROF routine by the means of a \(\chi^{2} \) fit (Napiwotzki 1999). The results of the fits with different grids were consistent (see Table 1).

\begin{center}
\vbox{\norm
\tabcolsep=13pt
\begin{tabular}{lccc}
\multicolumn{4}{c}{\parbox{90mm}{
{\bf \ \ Table 1.}{\ Parameters of KPD 1930+2752}}}\\
\tablerule
\multicolumn{1}{c}{model}&
\multicolumn{1}{c}{$T_{\rm eff}$}&
\multicolumn{1}{c}{$\log g$}&
\multicolumn{1}{c}{$\log \frac{n({\rm He})}{n({\rm H})}$}\\
\tablerule
LTE \, \(\log m = 0\)   & 35\,549 {\rm K}    & 5.63  &   -1.58\\
LTE \, \(\log m = +1\)   & 35\,183 {\rm K}    & 5.56  &  -1.48\\
NLTE \, (PRO2)   	& 35\,619 {\rm K}    & 5.56  &   -1.51\\
NLTE \, (NGRT)   	& 35\,516 {\rm K}    & 5.53  &   -1.45\\
NLTE \, (lf) & 35\,617 {\rm K}    & 5.64  &   -1.53\\
\tablerule
adopted & 35\,500 $\pm$ 500 ${\rm K}$  & 5.56 $\pm$ 0.05  & -1.48 $\pm$ 0.02\\
\tablerule
\end{tabular}
}
\end{center}

NLTE effects played a minor role. In the first instance it was not possible to fit the hydrogen and some helium lines (He\,{\sc i} 5876, He\,{\sc ii} 5412 and in particular He\,{\sc ii} 4686) simultaneously. This so called helium problem already occurred during other analyses of pulsating subdwarfs (Heber et al. 2000; Edelmann 2003). The analysis of HST-UV spectra of three sdB stars with similar \(T_{\rm eff}\) as KPD 1930+2752 revealed supersolar abundances of the iron group elements (O'Toole \& Heber 2005). Those stars also displayed the optical He ionisation problem. Using more appropriate metal-rich models (10 \(\times\) solar metalicity) the problem could be remedied (see also Heber et al. 2006). The abundances of the iron group elements have not been measured for KPD 1930+2752. Because of the similarity of its atmospheric parameters to those of the stars studied by O'Toole and Heber (2005), we adopted high-metalicity models as well and, indeed, the fit improved (Fig. 1).The derived temperature is about \(2\,000 \, {\rm K}\) higher than in prior analyses (Bill\`{e}res et al. 2000) irrespective of the choice of model atmosphere. With this KPD 1930+2752 is situated at the edge of the
instability strip in the \(T_{\rm eff} - \log \, g\) diagram (Charpinet et al. 1996).

\sectionb{4}{SPECTROSCOPIC ORBIT}

Due to the low signal to noise ratio of the HIRES spectra only the H$\alpha$ and H$\beta$ lines could be used for determining the radial velocity by \(\chi^{2}\) cross correlation with a model spectrum at rest wavelength. To improve the accuracy, the resulting radial velocity curve was combined with all available radial velocity data of KPD 1930+2752 (Maxted et al. 2000; Woolf et al. 2002; Orosz 2000 priv. comm.) covering a timespan of four years. A sine curve was fitted to these 2\,900 data points using a \(\chi^{2}\) minimizing method and the power spectrum was generated (FITRV and FITPOW routines by H. Drechsel). The sine curve fit is excellent and no period change could be detected over the whole timebase (Figs. 2). The orbital parameters were measured with unprecedented accuracy: \(\gamma({\rm H}\alpha)=5 \pm 1 {\rm \,km\,s^{-1}}\), \(K=341 \pm 1 \,{\rm km\,s^{-1}}\), \(P=0.0950933 \pm 0.0000015~ {\rm d}\). No radial velocity variations due to pulsations could be found.

\begin{figure}[t!]
\center\psfig{figure=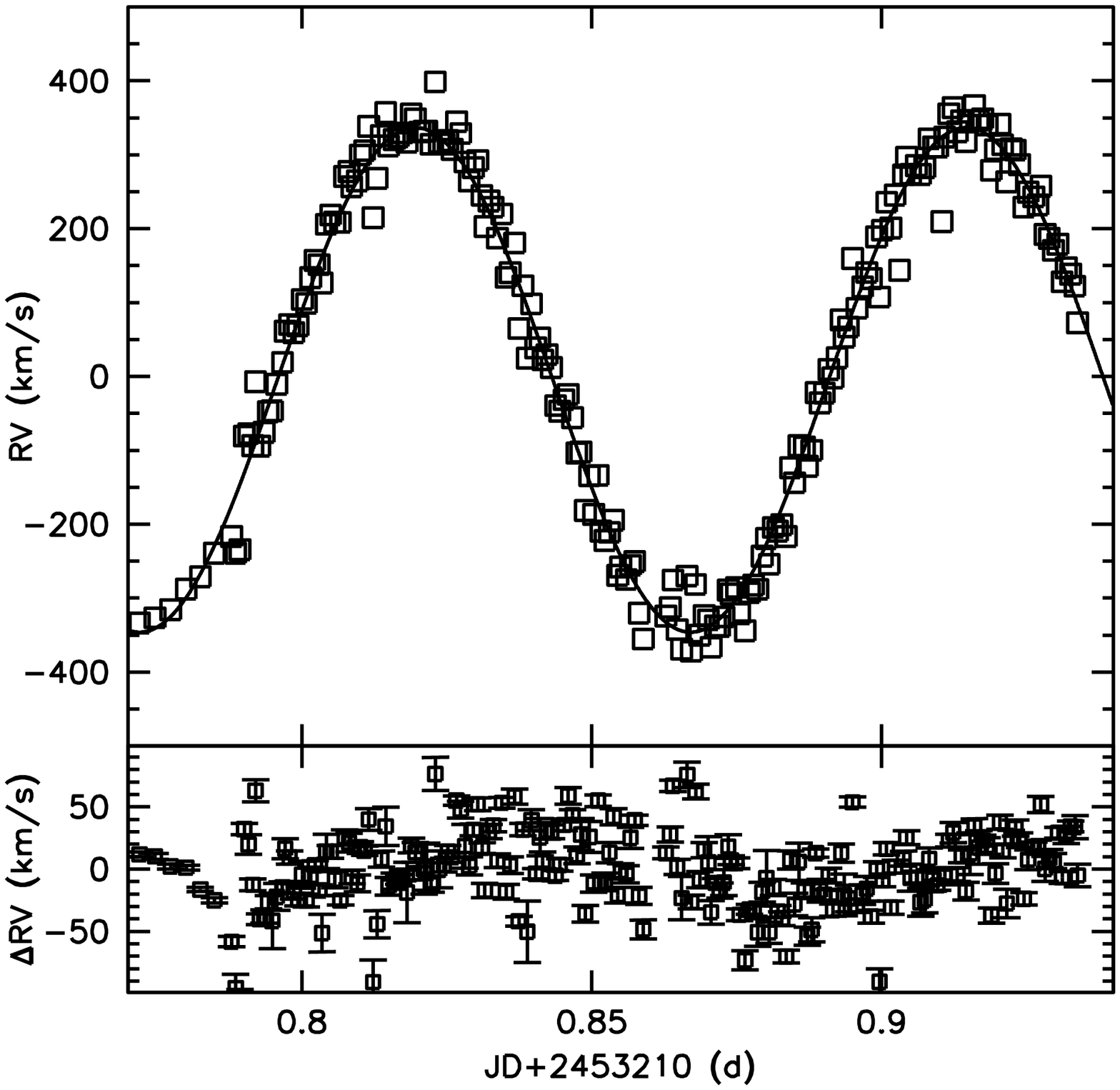,width=80mm,clip=}
\quote
\captionb{2}{Radial velocity curve. The data points are from the new HIRES spectra, the solid line is from the fit of all data points over a timespan of four years. The residuals in the lower part show no signs of eccentricity.}
\end{figure}

\begin{figure}[b!]
\center\psfig{figure=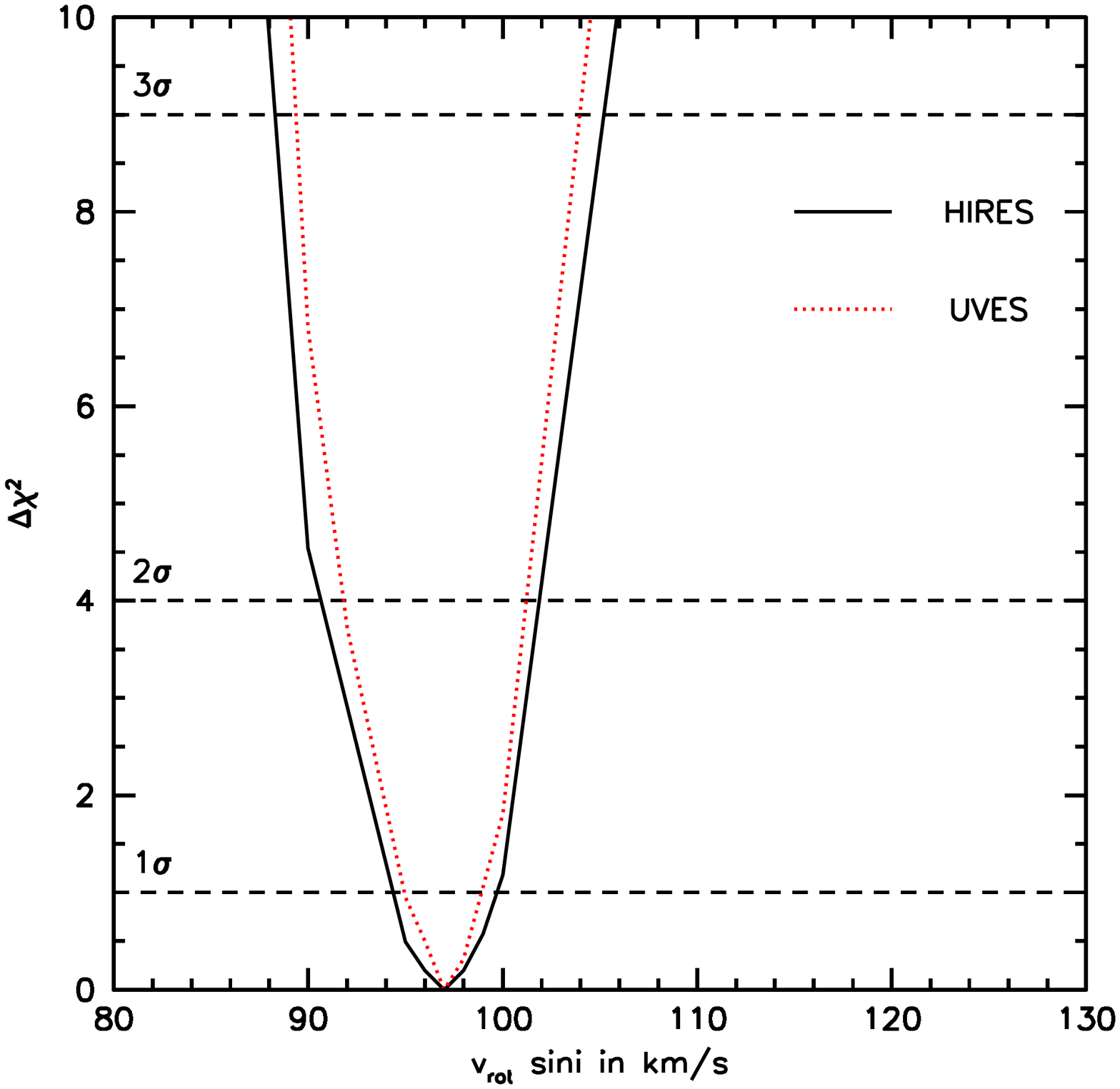,width=80mm,clip=}
\quote
\captionb{3}{Reduced $\chi^{2}$ as a function of $v_{rot} \sin{i}$ from fits to the HIRES and UVES data. The minimum of the two curves indicates the most probable value. The dashed horizontal lines mark the confidence levels of the errors.}
\end{figure}

\sectionb{5}{PROJECTED ROTATIONAL VELOCITY}

The main aim of the high-resolution time-series spectroscopy was to obtain a high-precision measurement of the projected rotational velocity.
For this purpose the 200 HIRES spectra were shifted to rest wavelength and the median was calculated in order to filter cosmics. The projected rotational velocity was measured using a model spectrum with the derived atmospheric parameters fixed and performing a \(\chi^{2}\) fit. The lines He\,{\sc ii} 4686, He\,{\sc i} 4922 and He\,{\sc i} 5016 were used for this measurement. Furthermore one single high resolution spectrum obtained with the ESO Very Large Telescope (VLT) at the Paranal Observatory and the Ultraviolet and Visual Spectrograph (UVES) was available. The rotational velocity was measured in the same way for the helium lines. The still preliminary results were consistent (see Fig. 3) and yield: \(v_{\rm rot}\sin{i} = 97 \pm 9 \, {\rm km\,s^{-1}}\, (3\sigma)\)

\sectionb{6}{MASS AND INCLINATION}

KPD 1930+2752 is obviously affected by the gravitional forces of the companion, demonstrated by its elipsoidal deformation.
Since the period of the photometric variations are exactly half the period of the radial velocity curve the rotation of the sdB star is very likely tidally locked to the orbit.
In addition to the mass function two more equations set additional constraints on the problem with only \(M_{\rm sdB}\) remaining as free parameter.

\begin{equation} \label{sini}
	\sin{i}=\frac{v_{\rm rot}\sin{i}P}{2\pi R} \quad ; \quad R=\sqrt{\frac{M_{\rm sdB}G}{g}}
\end{equation}

With \(\log \, g\), \(P\) and \(v_{\rm rot}\sin{i}\) measured and the fact that \(\sin{i}\) cannot exceed unity a lower limit for the mass of the sdB of \(0.44 \, M_{\odot}\) is derived. As can be seen in Fig.~4 the total mass of the system exceeds the Chandrasekhar limit for all reasonable assumptions of \(M_{\rm sdB}\). If the companion is a white dwarf, its mass has to be lower than the Chandrasekhar limit. This implies an upper limit \(M_{\rm sdB} \leq 0.63 \, M_{\odot} \) and a possible total mass range of \(M_{\rm sdB+WD} = 1.3 \pm 0.1 - 2.0 \pm 0.2 \, M_{\odot} \) (Fig. 5). Although a more massive companion cannot be ruled out completely, the object would probably be a bright X-ray source in this case. No such source is known at this coordinates.\\
The inclination angle of the system (Fig.~4) in the lower mass range is very close to \(90^{\circ} \). KPD 1930+2752 could be an eclipsing binary (cf. the very similar binary KPD 0422+5421, Orosz and Wade 1999), as some features in the light curve already indicated (Maxted et al. 2000).

\begin{figure}[h!]
\center\psfig{figure=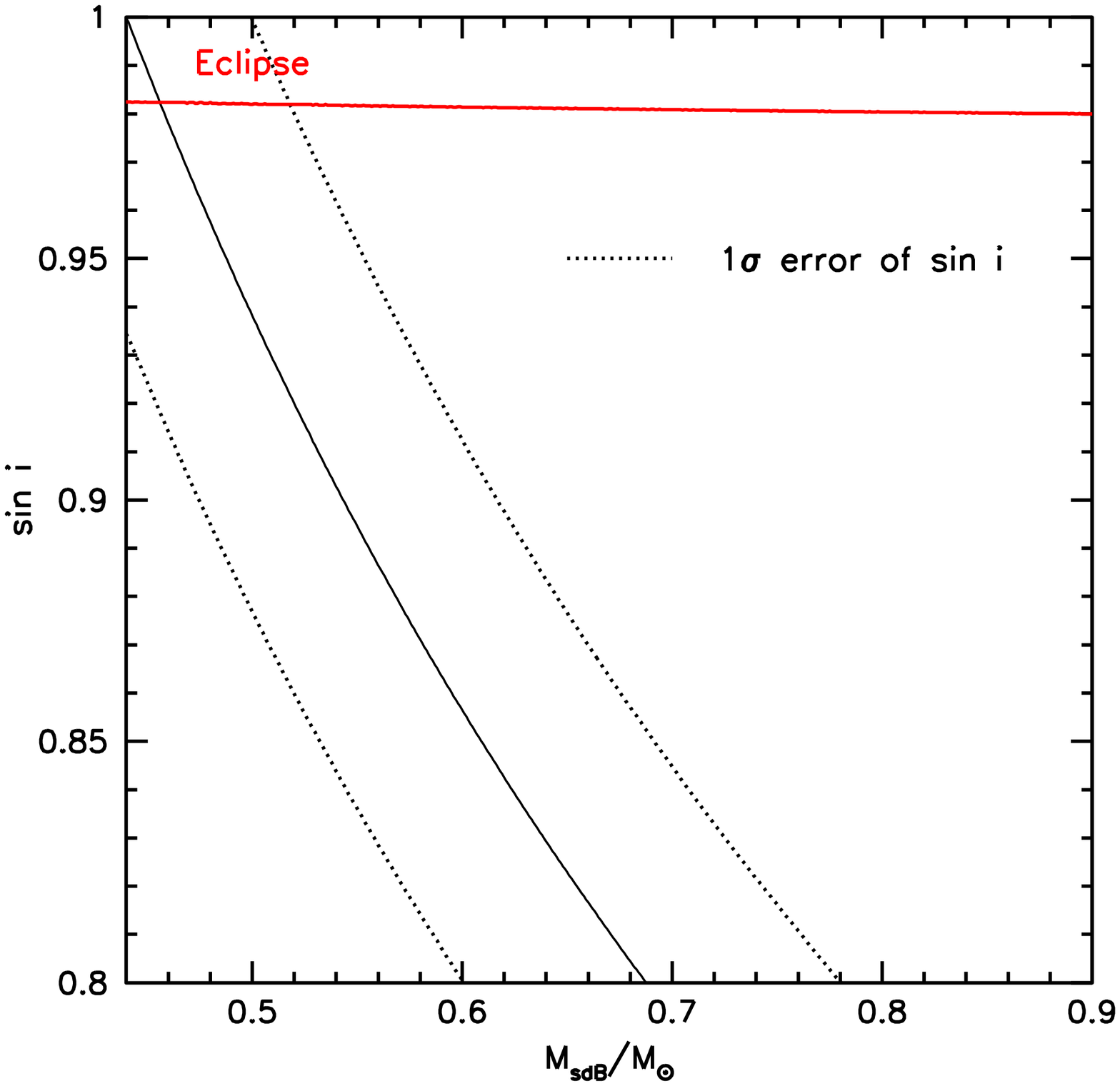,width=81mm,clip=}
\quote
\captionb{4}{Inclination with \(1 \sigma\) error (dotted line) versus mass of the sdB. The upper curve indicates the minimum inclination for an eclipse.}
\end{figure}

\begin{figure}[h!]
\center\psfig{figure=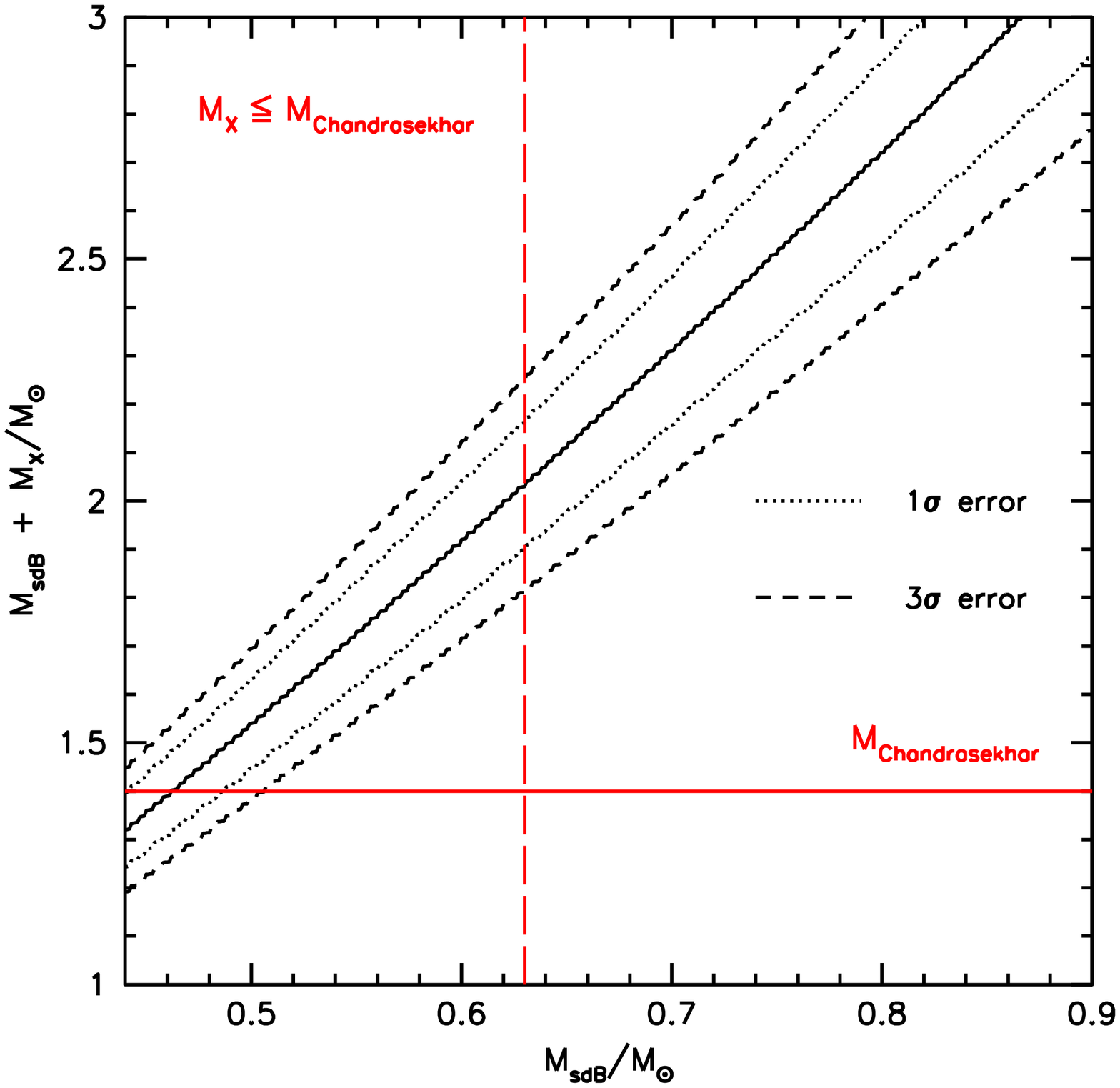,width=81mm,clip=}
\quote
\captionb{5}{Total mass of the binary as a function of the sdB mass. The dashed vertical line marks the point where the companion mass equals the Chandrasekhar mass.}
\end{figure}

\sectionb{7}{CONCLUSION}

The combination of an extensive set of 350 spectra and 2\,700 radial velocity data points from prior observations made it possible to determine the parameters of KPD 1930+2752 with very high accuracy. The tidally locked rotation together with a white dwarf companion implies a mass range for the sdB without any evolutionary model considerations. KPD 1930+2752 is the first subdwarf, whose mass could be constrained in this way. The total mass and the merging time of the binary indicate that it is a very good candidate for a SN Ia progenitor. KPD 1930+2752 is one of only three known double degenerate systems, which fulfill these requirements (Napiwotzki et al. 2003; Karl 2004).
While it fits into the DD scenario, KPD 1930+2752 might also evolve into a SN Ia via the SD scenario. Since time for merging due to gravitational wave radiation is of the same order as the EHB life time, Roche lobe overflow could occur well before the sdB becomes a white dwarf due to the shrinkage of the orbit. Recent calculations by Han and Podsiadlowski (priv. comm.) indicate that the mass transfer would be stable. Follow-up observations should be undertaken to measure an improved lightcurve and search for signs of an eclipse. If detected all system parameters will be determined. An alternative promising option is asteroseismology, which could tightly constrain the sdB mass.

\References

\refb
Bill\`{e}res M., Fontaine G., Brassard P., Charpinet S., Liebert J., Saffer, R. A., 2000, ApJ, 530, 441

\refb
Charpinet S., Fontaine G., Brassard P., Dorman B., 1996, ApJ, 471, L103

\refb
Downes R. A., 1986, ApJS, 61, 569

\refb
Dreizler S., Heber U., Werner K., Moehler S., de Boer K.~S., 1990, A\%A, 235, 234

\refb
Edelmann H., 2003, PhD-thesis, Friedrich Alexander Universit\"at Erlangen-N\"urnberg

\refb
Eggleton P.~P., 1982, ApJ, 268, 368

\refb
Ergma E., Fedorova A.~V., Yungelson L.~R., 2001, A\&A, 376, L9

\refb
Han Z., Podsiadlowski P., Maxted P.~F.~L., Marsh T.~R., Ivanova N., 2002, MNRAS, 336, 449

\refb
Han Z., Podsiadlowski P., Maxted P.~F.~L., Marsh T.~R., Ivanova N., 2003, MNRAS, 341, 669

\refb
Heber U., 1986, A\&A, 155, 33

\refb
Heber U., Reid I.~N., Werner K., 2000, A\&A, 363, 198

\refb
Heber U., et al., 2006, in Proc. of the Second Meeting on Hot Subdwarf Stars, Baltic Astronomy

\refb
Karl C., 2004, PhD-thesis, Friedrich Alexander Universit\"at Erlangen-N\"urnberg

\refb
Kilkenny D., Koen C., O'Donoghue D., Stobie R.~S., 1997, MNRAS, 285, 640

\refb
Koen C., Kilkenny D., O'Donoghue D., Van Wyk F., Stobie R.~S., 1997, MNRAS, 285, 645

\refb
Livio M., 2000, in Type Ia Supernovae: Theory and Cosmology, Cambridge Univ. Press, ed. Niemeyer J.~C., Truran J.~W., 33

\refb
Maxted P.~F.~L., Marsh T.~R., North R.~C., 2000, MNRAS, 317, L41

\refb
Napiwotzki R., 1999, A\&A, 350, 101

\refb
Napiwotzki R., Christlieb N., Drechsel H., et al., 2003, ESO Msngr, 112, 25

\refb
Orosz J.~A., Wade R. A., 1999, MNRAS, 310, 773

\refb
O'Toole S., Heber U., 2005, A\&A, submitted

\refb
Przybilla, N., Nieva M.~F., Edelmann H., 2006, in Proc. of the Second Meeting on Hot Subdwarf Stars, Baltic Astronomy

\refb
Saffer R.~A., Liebert J., 1995, in Proc. 9th European Workshop on White Dwarfs, ed. Koester D., Werner K., Springer Verlag, 221

\refb
Stobie R.~S., Kawaler S.~D., Kilkenny D., O'Donoghue D., Koen C., 1997, MNRAS, 285, 651

\refb
Vogt S., et al., 1994, Proc SPIE, 2198, 362

\refb
Woolf V.~M., Jeffery C.~S., Pollacco D.~L., 2002, MNRAS, 332, 34

\end{document}